\documentclass{ws-procs9x6}
\newcommand\beq{\begin{eqnarray}}
\newcommand\eeq{\end{eqnarray}}

\begin{document}

\title{$Q_T$-Resummation for Polarized Semi-Inclusive Deep 
Inelastic Scattering
}
\author{Yuji Koike and Junji Nagashima}
\address{Department of Physics, Niigata University, Ikarashi, 
Niiigata 950-2181, Japan\\
Email: koike@nt.sc.niigata-u.ac.jp}

\author{Werner Vogelsang}
\address{Physics Department,
Brookhaven National Laboratory, Upton, NY 11973, USA\\
E-mail: vogelsan@quark.phy.bnl.gov}  

\maketitle
\abstracts{
We study the transverse-momentum distribution of hadrons
produced in semi-inclusive deep-inelastic scattering. We
consider cross sections for various combinations of the
polarizations of the initial lepton and nucleon or the produced
hadron, for which we perform the resummation of large double-logarithmic
perturbative corrections arising at small transverse momentum. We present
phenomenological results for the process $ep\to e\pi X$ for the typical 
kinematics in the COMPASS experiment. We discuss the
impact of the perturbative resummation and of estimated non-perturbative
contributions on the corresponding cross sections and their spin
asymmetry.}

Semi-inclusive deep inelastic scattering (SIDIS)
with polarized beams and target, $ep\to ehX$, for which a hadron $h$ is 
detected in the final state, has been a powerful tool for
investigating the spin structure of the nucleon.  It also challenges
our understanding of the reaction mechanisms in QCD.
The bulk of the SIDIS events provided by experiments are in a 
kinematic regime of large virtuality $Q^2$ of the exchanged 
virtual photon and relatively small transverse momentum $q_T$.  
In our recent paper\,\cite{KNV06}, we have studied the
transverse-momentum dependence of SIDIS observables in this region,
applying the resummation technique of \cite{CSS85}.
The processes we considered were the leading-twist double-spin reactions: 
\beq
&{\rm (i)}&\quad e+p\rightarrow e+\pi+X,\qquad\qquad
{\rm (iv)}\quad \vec{e}+ \vec{p}\rightarrow e+ \pi+X \; ,\nonumber\\
&{\rm (ii)}&\quad e+\vec{p}\rightarrow e+ \vec{\Lambda}+X,\qquad\qquad\,
{\rm (v)}\quad \vec{e}+ p\rightarrow e+ \vec{\Lambda} +X \; .\nonumber\\
&{\rm (iii)}&
\quad e+ {p}^\uparrow\rightarrow e+ {\Lambda}^\uparrow +X \;,
\label{eq1.1}
\eeq
Here arrows to the right (upward arrows) denote longitudinal (transverse)
polarization. Needless to say, the final-state pion could be replaced
by any hadron. The same is true for the $\Lambda$, as long as the observed
hadron is spin-1/2 and its polarization can be detected experimentally.
Here we present a brief summary of the main results of \cite{KNV06}.

There are five Lorentz invariants for SIDIS,
$\vec{e}(k)+A(p_A,S_A) \rightarrow e(k') + B(p_B,S_B)+X$:
the center-of-mass energy squared for the initial electron and the proton, 
$S_{ep}=(p_A+k)^2$, the conventional DIS variables,
$x_{bj}={Q^2\over 2p_A\cdot q}$ and $Q^2 =-q^2=-(k-k')^2$, 
the scaling variable
$z_f={p_A\cdot p_B\over p_A\cdot q}$, and 
the magnitude of the ``transverse''  momentum 
$q_T = \sqrt{-q_t^2}$ where 
the space-like vector $q_t^\mu$ is defined as
$q_t^\mu=q^\mu- {p_B\cdot q\over p_A\cdot p_B}p_A^\mu -
{p_A\cdot q\over p_A\cdot p_B}p_B^\mu$ which is
orthogonal to
both $p_A$ and $p_B$.
To write down the cross section, we use
a frame where $\vec{p}_A$ and $\vec{q}$ are collinear, 
and we call the azimuthal angle between the
lepton plane and the hadron plane $\phi$.
In this frame, the transverse momentum of the final-state hadron $B$ 
with respect to $\vec{p}_A$ and $\vec{q}$ is
given by $p_T=z_fq_T$.

The lowest-order (LO) cross section differential in $q_T$ (or $p_T$)
is of $O(\alpha_s)$ and has been derived in \cite{KN03}.
It can be decomposed into several pieces with
different dependences on $\phi$:
\beq
{d^5\sigma\over dQ^2dx_{bj}dz_fdq_T^2 d\phi}
=\sigma_0 + {\rm cos}(\phi)\sigma_1 +
{\rm cos}(2\phi)\sigma_2 \; ,
\label{eq3unpol}
\eeq
for processes (i) and (ii) in~(\ref{eq1.1}),
\beq
{d^5\sigma\over dQ^2dx_{bj}dz_fdq_T^2 d\phi}
=\sigma_0 + {\rm cos}(\phi)\sigma_1 \; ,
\label{eq3pol}
\eeq
for (iv) and (v), and
\beq
\hspace*{-6mm}
& &{d^5\sigma_T\over dQ^2dx_{bj}dz_fdq_T^2 d\phi}
=\cos(\Phi_A-\Phi_B-2\phi)\sigma_0^T + \cos(\Phi_A-\Phi_B-\phi)
\sigma_1^T\nonumber\\
& &\qquad\qquad\qquad\qquad\qquad+\cos(\Phi_A-\Phi_B)\sigma_2^T \; ,  
\label{eq3Tpol}
\eeq
for (iii). Here $\Phi_A$ ($\Phi_B$) is the azimuthal angle of the
transverse spin vector of $A$ ($B$) as measured from the hadron plane around 
$\vec{p}_A$ ($\vec{p}_B$) in the so-called {\it hadron frame} 
for which $q=(0,0,0,-Q)$. 
At small $q_T$, $\sigma_0$ and $\sigma_0^T$ develop the large logarithmic
contribution $\alpha_s\ln(Q^2/q_T^2)/q_T^2$.  
At yet higher orders, corrections as large as 
$\alpha_s^k\ln^{2k}(Q^2/q_T^2)/q_T^2$ arise in the cross section.
We have worked out the NLL resummation of these large logarithmic corrections
in $\sigma_0$ and $\sigma_0^T$ for all the processes in (\ref{eq1.1})
within the $b$-space resummation formalism of\,\cite{CSS85}, 
extending the previous 
studies on the resummation for unpolarized SIDIS\,\cite{MOS96}.
The $\phi$-dependent contributions to the cross sections in general 
also develop large logarithms \cite{BV}; 
their resummation would require an extension of the formalism.

In order to study the impact of resummation, we have 
carried out a numerical calculation 
for the process $\vec{e}\vec{p}\to e\pi X$.  The resummed cross section 
takes the form of an inverse Fourier transform into $q_T$ space.
To carry out the Fourier integral, one needs a recipe for treating
the Landau pole present in the perturbatively calculated Sudakov form factor.  
We have followed the method of \cite{KSV02} which deforms the
$b$-integral to a contour integral in the complex $b$-plane. 
This method introduces no new parameter and is identical to
the original $b$-integral for any finite-order expansion of 
the Sudakov exponent.
For comparison, we have also used the $b^*$-method proposed in \cite{CSS85}.
In order to incorporate possible nonperturbative corrections,
we introduce a Gaussian form factor by shifting the Sudakov exponent as 
${\rm e}^{S(b,Q)}\to {\rm e}^{S(b,Q)-gb^2}$, where the 
coefficient $g$ may be determined by comparison with data. 
In order to obtain an adequate description also
at large $q_T\sim Q$, we ``match'' the resummed cross section
to the fixed-order (LO, $O(\alpha_s)$) one.
This is achieved by subtracting from the resummed expression its 
$O(\alpha_s)$ expansion
and then adding the full $O(\alpha_s)$ cross section\,\cite{KSV02,BCDG03}.

As an example, we show in Fig. 1 the $z_f$-integrated cross sections
\beq
{1\over 2\pi}\int_{0.2}^{z_f^{max}}dz_fd\phi{d(\Delta)\sigma
\over dx_{bj} dz_f dQ^2 dq_T d\phi},
\eeq
and their spin asymmetry for the typical kinematics of the COMPASS experiment,
$S_{ep}=300$ GeV$^2$, $Q^2=10$ GeV$^2$, $x_{bj}=0.04$.  As expected,
the resummation tames the divergence of the LO cross section
at $q_T\to 0$ and enhances the cross section in the region of 
intermediate and large $q_T$.
The nonperturbative Gaussian makes this tendency stronger. 
Although the cross sections vary slightly when different
treatments of the $b$-integral and different values of $g$ are chosen, 
the effects of resummation and the nonperturbative Gaussian
are mostly common to both the unpolarized and the polarized cases.
Accordingly, the spin asymmetry is relatively insensitive to these effects.
It will be interesting to compare our results with forthcoming 
data from COMPASS and HERMES, and also to extend the analysis
to the reaction $\vec{e}p\to e\vec{\Lambda}X$ which is accessible
at HERA.

\begin{figure}[ht]
\hbox{\epsfxsize=2.4in\epsfbox{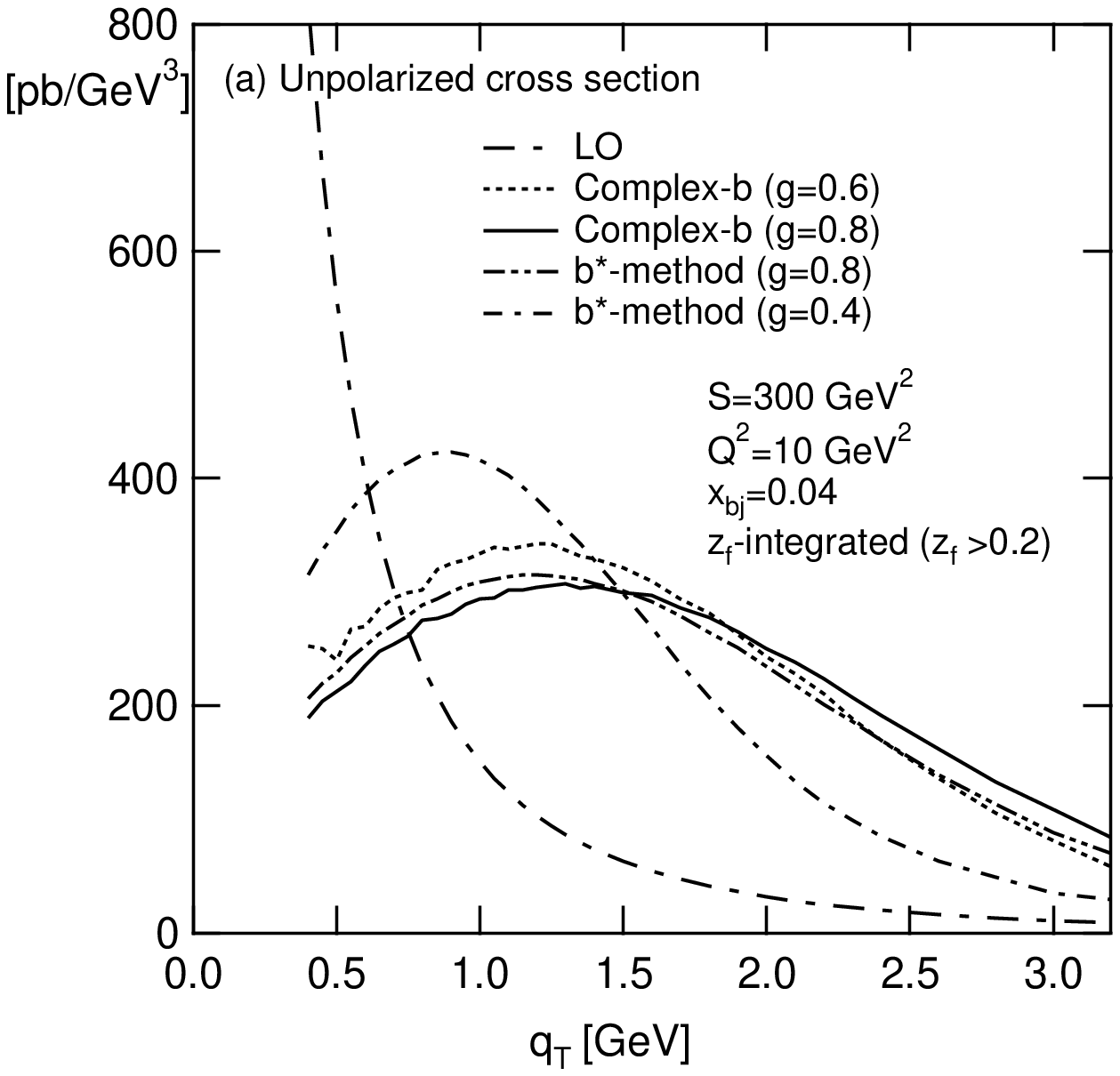}\hspace{-0.4cm}
\epsfxsize=2.4in\epsfbox{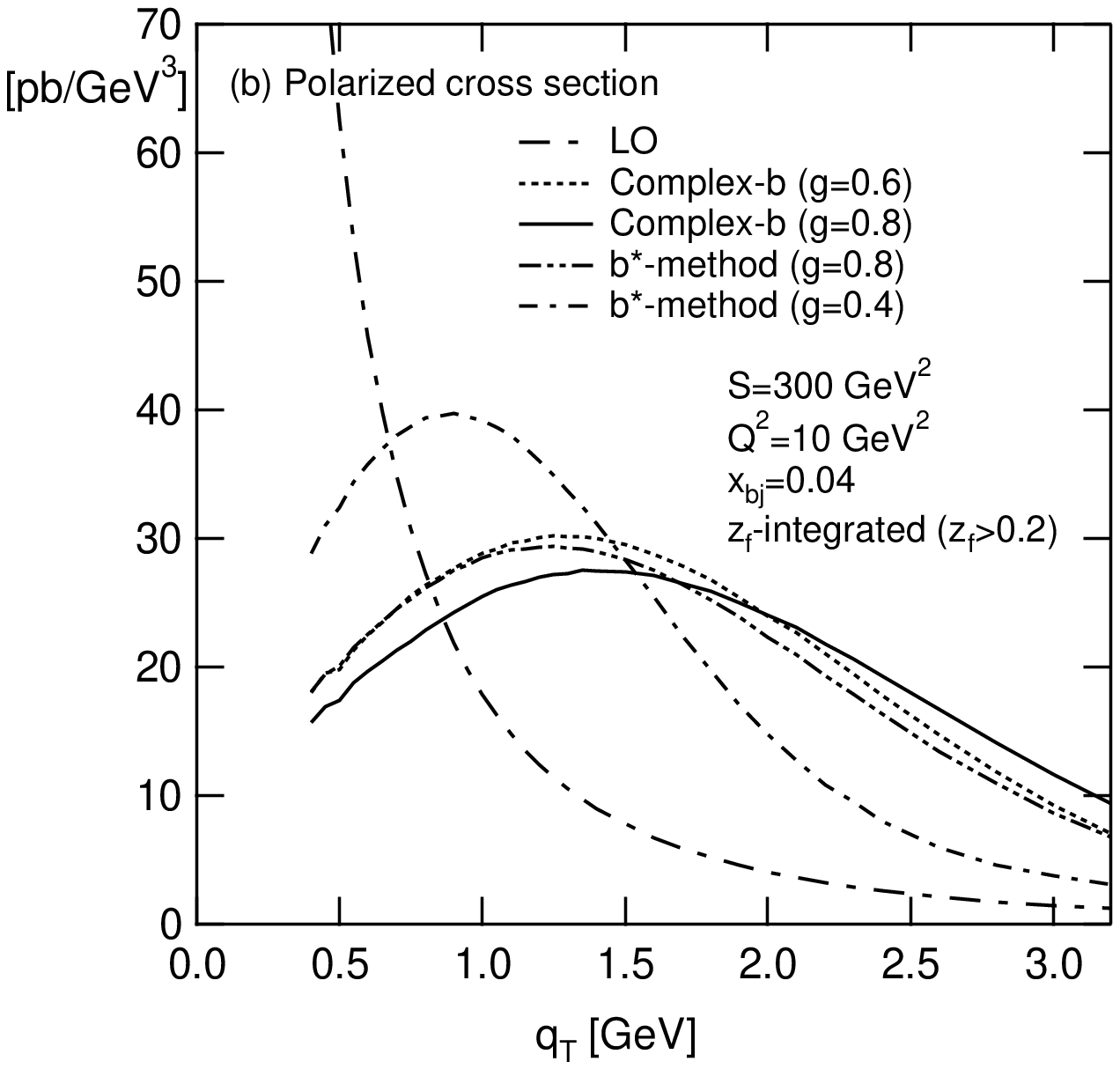}}
\centerline{\epsfxsize=2.4in\epsfbox{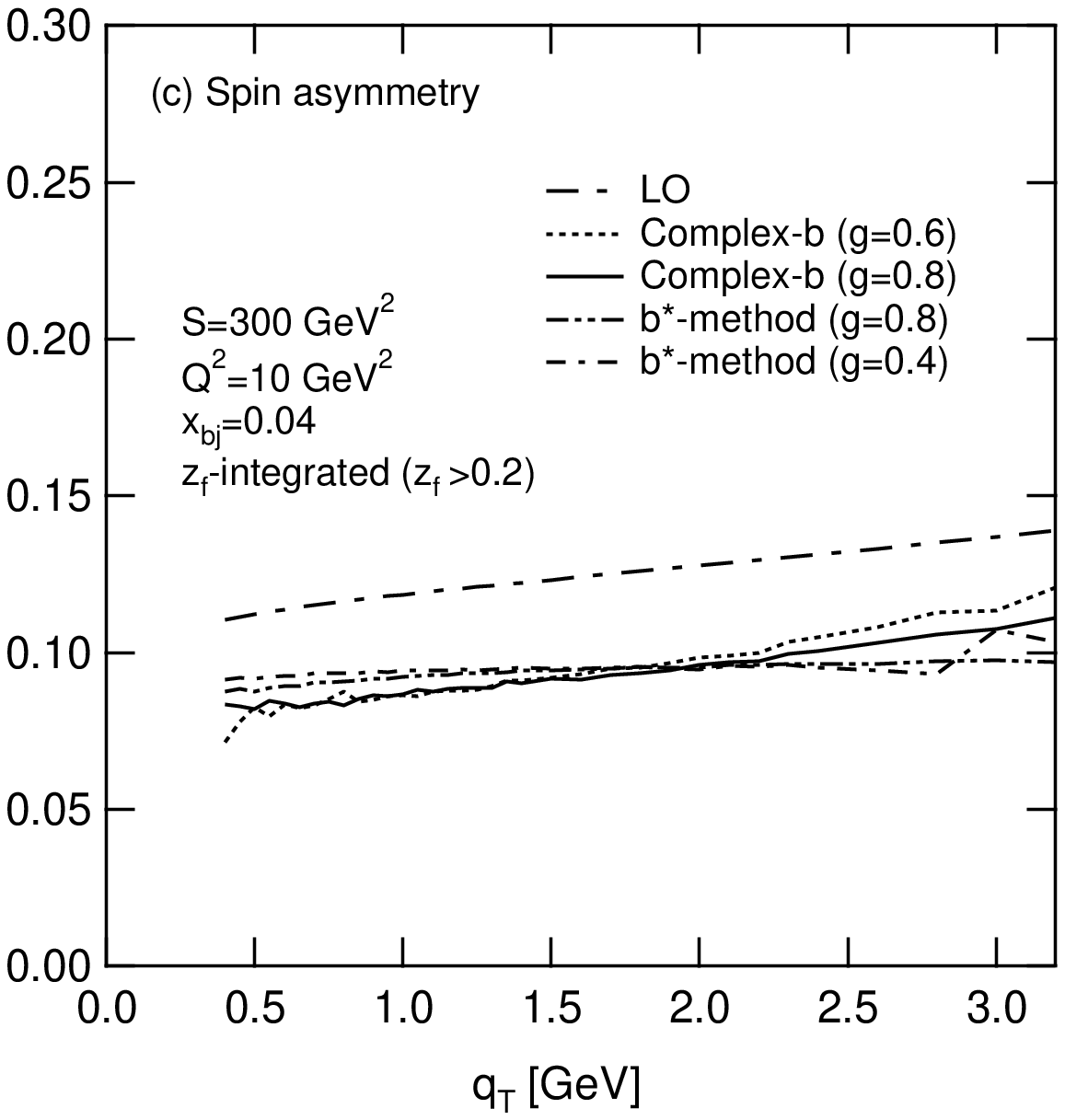}}
\caption{
(a) Unpolarized SIDIS cross section for COMPASS kinematics.
We show the fixed-order (LO) result, and resummed results for the
complex-$b$ method with non-perturbative parameters 
$g=0.6,\,0.8$~GeV$^2$, and for the $b^{\ast}$
method with $b_{\rm max}=1/(\sqrt{2}$~GeV$)$ and $g=0.4, \, 0.8$~GeV$^2$.
(b) Same for the longitudinally polarized case. (c) Spin asymmetries
corresponding to the various cross sections shown in~(a) and~(b).
\label{figure1}}
\end{figure}

\section*{Acknowledgments}
W.V. is supported by DOE Contract No. DE-AC02-98CH10886.

\end{document}